# Probing the Kondo Lattice

Satoru Nakatsuji[*§], David Pines[†‡], and Zachary Fisk[*†]

[*]National High Magnetic Field Laboratory, Florida State University, Tallahassee, FL 32310,

[†] K764 Los Alamos National Laboratory, Los Alamos, NM 87544, [‡] Institute For Complex Adaptive Matter, University of California.

[§] Present address: Department of Physics, Kyoto University, Kyoto 606-8502, Japan.

**Remarkably rich physics is found in the vicinity of the magnetic/non-magnetic boundary in matter, a region that includes the high $T_c$ superconductors, giant magnetoresistance materials, and especially the heavy electron superconductors [1]. The theoretical framework for the physics of heavy electron materials is the Kondo lattice that has, at high temperatures, localized electronic magnetic moments at each lattice site weakly coupled antiferromagnetically to an itinerant electron band (Fig. 1). The single site impurity problem for the coupled electron has been solved to describe the Kondo effect that develops at low temperatures [2,3]. Yet for the Kondo lattice, because of complex intersite coupling effects, the general form of a solution has not been found despite over thirty years of experimental and theoretical effort. We present here a solution of surprising simplicity that is derived from an analysis of the bulk specific heat and spin susceptibility of the heavy electron superconductor $CeCoIn_5$, a particularly interesting member of the 1-1-5 family ($CeMIn_5$) of Kondo lattice materials that exhibits an unconventional Cooper pairing with $T_c$ of 2.3 K and non-Fermi-liquid behavior due to its proximity to an antiferromagnetic instability [4-9]. We**

**find that below a crossover temperature corresponding to the intersite coupling scale, $T^* \sim 45$ K, the Kondo gas (of non-interacting Kondo impurities) partially condenses into a heavy electron "Kondo" liquid that has a temperature independent Wilson ratio = 2.0. The relative fraction, $f$, of the "condensed" Kondo liquid component plays the role of an order parameter; it increases linearly with decreasing temperature until it saturates at its low temperature value of 0.9. The resistivity is shown to be simply the product of $(1-f)$ and that of an isolated Kondo impurity. The generality of this result is suggested by the corresponding analysis for $Ce_{1-x}La_xCoIn_5$ and $CeIrIn_5$.**

We begin our analysis at low temperatures by decomposing $C_{MAG}$, the magnetic f-electron part of the specific heat of $Ce_{1-x}La_xCoIn_5$ alloys (obtained by subtracting the background lattice specific heat, that of $LaCoIn_5$) into two components,

$$C_{MAG}/T = [1 - f(T)](C_{KG}/T) + f(T)(C_{KL}/T) \qquad (1)$$

where $C_{KG}$ is the single impurity contribution of the f-electrons part, $C_{KL}$ is the itinerant heavy electron, or Kondo liquid contribution, and $f(T)$ is its relative weight. In Fig. 2a we plot $C_{MAG}/T$, normalized per Ce, versus $C_{KG}/T$ of isolated Kondo impurity sites obtained from the measured specific heat of $Ce_{1-x}La_xCoIn_5$ alloys at low Ce concentrations ($x > 0.95$). $C_{KG}$ agrees very well with the exact results of $S = 1/2$ Kondo impurity model with the single parameter $T_K = 1.7$ K [3,9]. In this plot, $T$ is an implicit variable and low $T$ corresponds to large values of $C_{MAG}/T$. By definition, the slope is 1 for the single impurity limit with $x$

= 0.97, 0.98. Notably, even with increasing the Ce concentration, the linear relation between $C_{MAG}/T$ and $C_{KG}/T$ holds at low $T$ with a slope decreasing systematically from 1. For $0.25 > x \geq 0$, the slope converges to 0.1, indicating that the variation of $C_{MAG}/T$ at low $T$ is a linear function of the single impurity contribution $C_{KG}/T$: $C_{MAG}/T = A + BC_{KG}/T$ with $A \approx 290$ mJ/mol-Ce $K^2$ and $B \approx 0.1$. Fig. 2b shows that an analogous linear relation for the susceptibility, $\chi = C + D\chi_{KG}$, is found for the entire alloying range between the measured $\chi$ and $\chi_{KG}$ of isolated Kondo impurity sites, obtained at the single impurity limit. Again for $0.25 > x \geq 0$, the slope D converges to 0.1 with a constant $C \approx 0.008$ emu/mol-Ce. At low temperatures it thus appears that each f-electron of the dense Kondo lattice acts as though it were a mixture of 10% Kondo gas and 90% Kondo liquid. We note that CeCoIn$_5$ is a particularly clean material for analysis, having the three important energy scales of the Kondo lattice, $T_K$ (1.7 K), $T^*$ (45 K) and crystal electric field (CEF) splitting (120 K) well separated, while the La dilution keeps essentially the same $T_K$ and CEF parameters [9]. We further find that the low temperature Wilson ratio for the Kondo liquid component, $R_W = \alpha\chi/(C/T)$, where $\alpha = (\pi^2 k_B^2/3\mu_B^2)$, is 2.0, a result identical to that for the Kondo gas component [2,3].

In general, $C_{KL}/T$, $C_{KG}/T$, and $f$ all have temperature dependence. $C_{KG}/T$ is known, from the experimentally determined specific heat of dilute Ce in LaCoIn$_5$, as is $C_{MAG}/T$ from the measurements on CeCoIn$_5$. To obtain the temperature dependence of both $f$ and $C_{KL}$, we first write the magnetic susceptibility in a form analogous to Eq. (1):

$$\chi(T) = [1 - f(T)] \chi_{KG}(T) + f(T) c_{KL}(T) \qquad (2)$$

and then make the ansatz (we refer to this as heavy electron ansatz below.) that for all temperatures below $T^*$ the Wilson ratio of the specific heat to the spin susceptibility of the Kondo liquid components of Eqs. (1) and (2) is always 2, i.e. $\chi_{KL}(T) = 2 \gamma_{KL}(T)/\alpha = 2 C_{KL}(T)/\alpha T$. We can then eliminate $\chi_{KL}$ from the two equations and determine $f(T)$ and $C_{KL}(T)/T$ from experiment, with the result shown in Fig. 3a and 3b. In Fig. 3a, we see that $f(T)$, which plays a role analogous to that of an order parameter for the gas-liquid crossover, increases linearly with $T$ as $T$ is reduced below $T^*$, becoming constant around $T = 2$ K. For CeCoIn$_5$, we find that $f(0) \sim 0.9$ and $T^* \sim 45$ K. Figure 3b shows the $T$ dependence of the decomposed liquid part $fC_{KL}/T$ and gas part $(1-f)C_{KG}/T$ determined uniquely by our heavy electron ansatz. We find that $C_{KL}(T)/T$, corresponding to an average effective mass, $m^*$, develops as $\ln T$ below $T^*$, as shown in the inset of Fig. 3b.

What is remarkable about the two-component description of f-electron behavior that emerges from our analysis is that both components exhibit protected behavior [10]; the Kondo gas component behaves like a "Wilson quantum protectorate" in which each f-electron is independently screened, while the heavy electron component exhibits yet another form of protected behavior, a temperature independent Wilson ratio and a logarithmic increase in effective mass that reflects the local origin of the hybridization of the f- and d-electrons responsible for the enhanced heavy-electron effective mass and spin susceptibility. The presence of these two distinct components could be argued to represent intrinsic spatial phase separation, but there is no direct experimental evidence to support such a view. We therefore attribute it to a redistribution of the spectral weight of the f-electrons. The local moment part is incoherent and refers to the high-frequency component of the f-electron spectral weight; the itinerant heavy electron part is coherent and represents the low frequency part. It would be Landau Fermi-liquid-like, were it not for the possible influence of quantum criticality or strong antiferromagnetic correlations. What begins below $T^*$ is then a transfer of spectral weight from an incoherent local high frequency part, whose behavior is protected by the Kondo result of Wilson, to the coherent itinerant low frequency part of the f-electron spectral weight. On this argument, the f-electron part of the Fermi surface should be that for all the f-electrons, rather than 90 % of them for $CeCoIn_5$.

There are three independent checks on our analysis of $CeCoIn_5$. First, if we make the ansatz that the magnetic part of the electrical resistivity of $CeCoIn_5$ (obtained by subtracting the temperature dependent resistivity of $LaCoIn_5$ from that of $CeCoIn_5$) comes solely from that part of electron–electron scattering that corresponds to scattering off the local Kondo centers, one can use our two component decomposition to calculate the resistivity of $CeCoIn_5$. We thus take the measured Ce single impurity resistivity in $LaCoIn_5$, scale this to 100 % Ce, and multiply by $1-f(T)$, which is the time averaged fraction of such centers. We plot without any adjustment this expression and compare with the measured curve in Fig. 4. Experiment is remarkably well reproduced; in particular the so-called coherence peak is now seen to due to the loss of local Kondo centers as the heavy electron Kondo liquid is formed, and the peeling away of the resistivity from the single impurity result reflects the onset of Kondo liquid formation at $T^*$. That the fit is not perfect may reflect many-body modifications in the scattering at low temperatures.

A second check is provided by calculating the magnetic field response of the magnetization $M$. While the Kondo liquid part should give a linear increase of $M$, the local moment part should generate the sum of a van Vleck linear part and a Brillouin-type non linear response due to the saturation of the local moments. The total magnetization is thus decomposed as: $M = fM_{KL} + (1-f)M_{KG}$, where $M_{KL} = \chi_{KL}H$ and $M_{KG} = g\mu_B\{2a(g\mu_BH) + b \tanh(ag\mu_BH/k_B(T+T_K))\}$. Here, a and b are constants for the CEF scheme obtained at the single impurity limit. We can then calculate the field dependence of $M$ without any tuning parameters. The inset of Fig. 4 shows that the experimental results from Ref. 11 are well reproduced by our calculation.

Third, we note that it is plausible that when the material goes superconducting, the local moments should become part of the superconducting condensate. The usual weak coupling BCS expression for the specific heat jump at $T_c$ is $\Delta C/\gamma T_c = 1.43$. Recognizing that $\gamma T_c$ is the electronic entropy at $T_c$, we can rewrite the BCS expression as $\Delta C/S(T_c) = 1.43$. Integrating the low temperature entropy out to $T_c = 2.3$ K for CeCoIn$_5$ gives $\Delta C/S(T_c) = 2.2$ with $S(T_c) \approx 0.2$ Rln2. Given that $S(T_c)$ should have a large contribution from the Kondo gas part of about $(1 - f(0))$Rln2 = 0.1 Rln2, this strong coupling value of $\Delta C/S(T_c)$ indicates that the superconducting ground state incorporates fully the local Kondo gas component into the superconducting condensate. This also shows that the local moment part is an intrinsic property of the material.

In order to test the generality of our scheme found for $CeCoIn_5$, we adopt a similar approach in analyzing the experimental results for $Ce_{1-x}La_xCoIn_5$ and $CeIrIn_5$. The inset of Fig. 2b shows $f(0)$ for $Ce_{1-x}La_xCoIn_5$ obtained from the linear slope of $C_{MAG}/T$ in Fig. 2a and $\chi$ in Fig. 2b. They both increase systematically from 0 with increasing the Ce content from the single impurity limit, converging to essentially the same values for the dense Kondo regime with $x \leq 0.5$. This indicates that it is the intersite coupling that drives the condensation of the Kondo gas. Interestingly, as shown in the inset of Fig. 3a, when $f(T)$ is normalized by $f(0)$, $f(T)/f(0)$, it scales with $T^*$; it appears around $T^*$ and linearly increases with decreasing $T$, indicating that $T^*$ should give the condensation energy scale (see the inset of Fig. 2b). Scaling behavior with $T^*$ is also found for $C_{KL}(T)/T$; $C_{KL}(T)/T = A/T^* - B/T^* \ln(T/T^*)$ with $A \approx 4.5$ J/mol-Ce K and $B \approx 3.3$ J/mol-Ce K above $T \sim 0.05T^*$, and $C_{KL}(T)/T = D/T^*$ with $D \approx 14.5$ J/mol-Ce K below $T \sim 0.05T^*$, as shown in the inset of Fig. 3b. These scaling functions possess the same form as that found for the single impurity Kondo problem with $T_K$ [2,3] and should be considered as yet another protected behavior of the Kondo lattice. The $T$ linear dependence of $f(T)$ and the $\ln T$ dependence of $C_{KL}(T)/T$ are so robust that they survive up to 50 % La dilution. In this range, $T^*$ increases linearly with the Ce concentration, as shown in the inset of Fig. 2b. Since other characteristic energy scales of the Kondo lattice, $T_K$ and the CEF splitting, are kept constant with La dilution [9], this linear increase in the condensation energy scale $T^*$ must arise from the intersite coupling. In fact, $T^*$ is basically the same as the energy scale for intersite interactions determined in Ref. 9.

Our analysis carries over to the isostructural heavy fermion superconductor, CeIrIn$_5$, with $T_c$ = 0.4 K [12]. We show the plot of $C_{MAG}$ vs. $C_{KG}$ in the inset to Fig. 2a. Again it is seen that the low temperature specific heat of the pure Ir material is linear in that of its dilute Ce counterpart, while $T_K$ and $T^*$ are found to be 2.7 K and 20 K respectively. Thus, the $T \sim 0$ normal ground state is well described by the sum of Kondo gas and Kondo liquid parts, the latter being here 0.95 of the whole. Furthermore, $f(T)$ and $C_{KL}(T)/T$ scale with $T^*$ at low temperature and collapse on top of the respective $T$ linear and ln$T$ behavior obtained for Ce$_{1-x}$La$_x$CoIn$_5$ system, as shown in the insets of Figs. 3a and 3b. Neutron experiments show that the next higher lying doublet of the $J = 5/2$ 5f Hund's rule multiplet lies at 70 K, compared to 120 K for CeCoIn$_5$ [13]. This difference may be the origin of the deviations seen for $(T/T^*) > 0.2$ for the Ir material. It is further worth noting that here again the Kondo gas remnant at the superconducting $T_c$ is incorporated into the superconducting condensate.

Our preliminary analysis of the antiferromagnet CeRhIn$_5$ ($T_N$ = 3.8 K)[14], finds that $f \approx 1$ at low temperatures. We describe this as the complete condensation of the Kondo gas into heavy electron Kondo liquid. The antiferromagnetism of this compound then is seen as a Fermi surface instability rather than a local moment instability, a surprising consequence of our analysis. The observation that the electronic specific heat goes to 0 as $T$ goes to 0 K [14], means that the magnetic order destroys the heavy fermion state or, put another way, recaptures the spin moments upon which the heavy electron Kondo liquid arises.

Our picture is one of remarkable simplicity. The Kondo lattice at high temperatures is a gas of non-interacting Kondo centers that start condensing into a heavy electron Kondo liquid below a characteristic temperature $T^*$ corresponding to the intersite coupling scale.

This process represents the gradual transfer, as the temperature is lowered, of the spectral weight for each f-electron from its high temperature, high frequency, incoherent localized value to the low frequencies expected for coherent itinerant heavy electron behavior. It is well represented by the universal $T$ linear increase of $f(T)$ and $\ln T$ increase of $C_{KL}(T)/T$ found for both $Ce_{1-x}La_xCoIn_5$ and $CeIrIn_5$ systems. Experiment suggests that these 1-1-5 materials are located close to an antiferromagnetic quantum critical point, namely near the point where an antiferromagnetic ordering temperature vanishes [7-9,12,14,15]. Our results indicate that, in the vicinity of the quantum critical point, this condensation appears to be incomplete; a local Kondo center contribution remains, a contribution that may be closely connected to superconductivity.

Our picture differs from the usual qualitative picture of heavy fermion materials in which the dense lattice of f-moments is characterized by a Kondo temperature appropriate to the lattice and distinct from the single impurity scale. We see here the survival of the single impurity scale side by side with the lattice scale characterizing the strength of the interaction between the Kondo centers that is responsible for the condensation of the Kondo gas. The "coherence" feature so prominent in the temperature dependent electrical resistivity of heavy fermions is found to come simply from the low temperature loss of scattering from the Kondo gas component.

Our phenomenology suggests the form that the low temperature solution to the Kondo lattice must take. It provides an unexpected framework within which to view the Kondo lattice and heavy electron behavior, and has the potential to provide a new unifying view of correlated electron materials more generally. It provides an experimental algorithm for analyzing heavy fermion systems of all kinds, and represents a new tool for investigating quantum critical behavior and the occurrence of superconductivity in highly correlated electron systems.

**Acknowledgments.**  This research originated in discussions at an ICAM workshop on 1-1-5 materials held in December, 2002 for which we thank our fellow participants. In addition we wish to acknowledge stimulating discussions with L.P. Gor'kov, P. Schlottmann. We also thank H. Lee, N. Curro, R. Movshovich, M. Hundley, N. Moreno, and J.D. Thompson for sharing their data with us. This work was supported in part by grant NSF DMR – 0203214 (S.N. and Z.F.) and NSF DMR – 0084540 (D.P.). Work at Los Alamos National Laboratory supported by U.S. Department of Energy. S.N. was supported by JSPS.

Correspondence should be addressed to D.P. (e-mail: pines@lanl.gov)

**Figure 1**, Schematic diagram of the Kondo lattice. For the single impurity case, a local moment (green arrow) couples with a spin (blue arrow) of conduction electron bands through an onsite antiferromagnetic interaction (solid line), whose screening energy scale of the local moment is represented by the Kondo temperature $T_K$. For the Kondo lattice, there is an additional intersite coupling (broken line) mediated by conduction electrons between the local moments at neighboring lattice sites $i$ and $j$. The characteristic temperature $T^*$ in the text gives its energy scale.

**Figure 2 a**, The f-electron contribution to the specific heat divided by temperature $C_{MAG}/T$ vs. its single impurity limit counterpart $C_{KG}/T$ for $Ce_{1-x}La_xCoIn_5$ with temperature as an implicit variable. Inset: The same plot for $Ce_{1-x}La_xIrIn_5$, from which we deduce that $f = 0.95$ for $CeIrIn_5$. The specific heat was measured by a relaxation method using high-quality single crystals grown by an In self-flux method [4]. **b**, $\chi$ vs. $\chi_{KG}$ for $Ce_{1-x}La_xCoIn_5$ with temperature as an implicit variable. Here, $\chi$ is the magnetization per mole Ce divided by the field along the c-axis of the single crystals, and $\chi_{KG}$ is the one for the single impurity limit. The magnetization was measured down to 1.8 K with a Quantum Design MPMS SQUID magnetometer. Inset: The left axis shows $f(0)$ derived from $C_{MAG}/T$ in Fig. 2a (red open symbol) and from $\chi$ in Fig. 2b (blue solid symbol) for $Ce_{1-x}La_xCoIn_5$. $T^*$ is given for the right axis. The broken line is a linear fit.

**Figure 3 a**, The relative weight of the Kondo lattice component $f(T)$ for CeCoIn$_5$ as a function of $T$. The broken line shows a linear fit. Inset: The scaling behavior with $T^*$ of $f(T)/f(0)$ for Ce$_{1-x}$La$_x$CoIn$_5$ and CeIrIn$_5$. The solid line represents a fit to the linear increase of $f(T)$ that universally appears below $\sim T^*$. **b**, The decomposition of the f-electron contribution to the normal state specific heat $C_{MAG}/T$ of CeCoIn$_5$ into Kondo gas $(1-f)C_{KG}/T$ and Kondo liquid components $fC_{KL}/T$. $C_{MAG}/T$ and $fC_{KL}/T$ above and below $T_c = 2.3$ K represent the data taken under fields of 0 T (solid symbol) and 5 T (open symbol) along the c-axis, respectively [4]. For $(1-f)C_{KG}/T$, solid and open symbols are the experimental data taken at the single impurity limit with $x = 0.97$ and 0.98, respectively. The solid curve is the fit to the exact results of $S = 1/2$ Kondo impurity model. Inset: $T^*C_{KL}/T$ as a function of $\ln(T/T^*)$ for Ce$_{1-x}$La$_x$CoIn$_5$ and CeIrIn$_5$ under $B = 0$ T. The solid line is a fit to the $\ln(T/T^*)$ behavior that universally appears for $0.05T^* < T < T^*$.

**Figure 4**, The magnetic part of the in-plane resistivity $\rho_{MAG}$ defined as the difference of the temperature dependence of the resistivity of CeCoIn$_5$ and LaCoIn$_5$. The black line is for CeCoIn$_5$, while the blue line corresponds to $\rho_{MAG}$ measured at the single impurity limit, $\rho_{KG}$. The red open circles are the calculated resistivity, $(1-f)\rho_{KG}$, for CeCoIn$_5$. Here, $f(T)$ is given in Fig. 3a with its linear extrapolation to 0 around $T^*$. The resistivity was measured by the standard four-probe dc technique down to 0.4 K. Inset: The field dependence of the c-axis magnetization of CeCoIn$_5$ at 50 mK. The calculated magnetization based on our two-component scheme (red open circle) is compared with the experimental results (black line) presented in Ref. 11. The agreement is nice in the normal state under $B > B_{C2} = 5$ T.

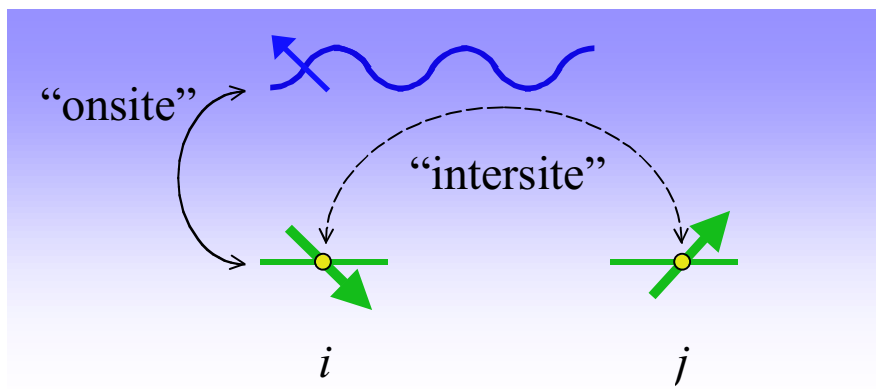

Figure 1

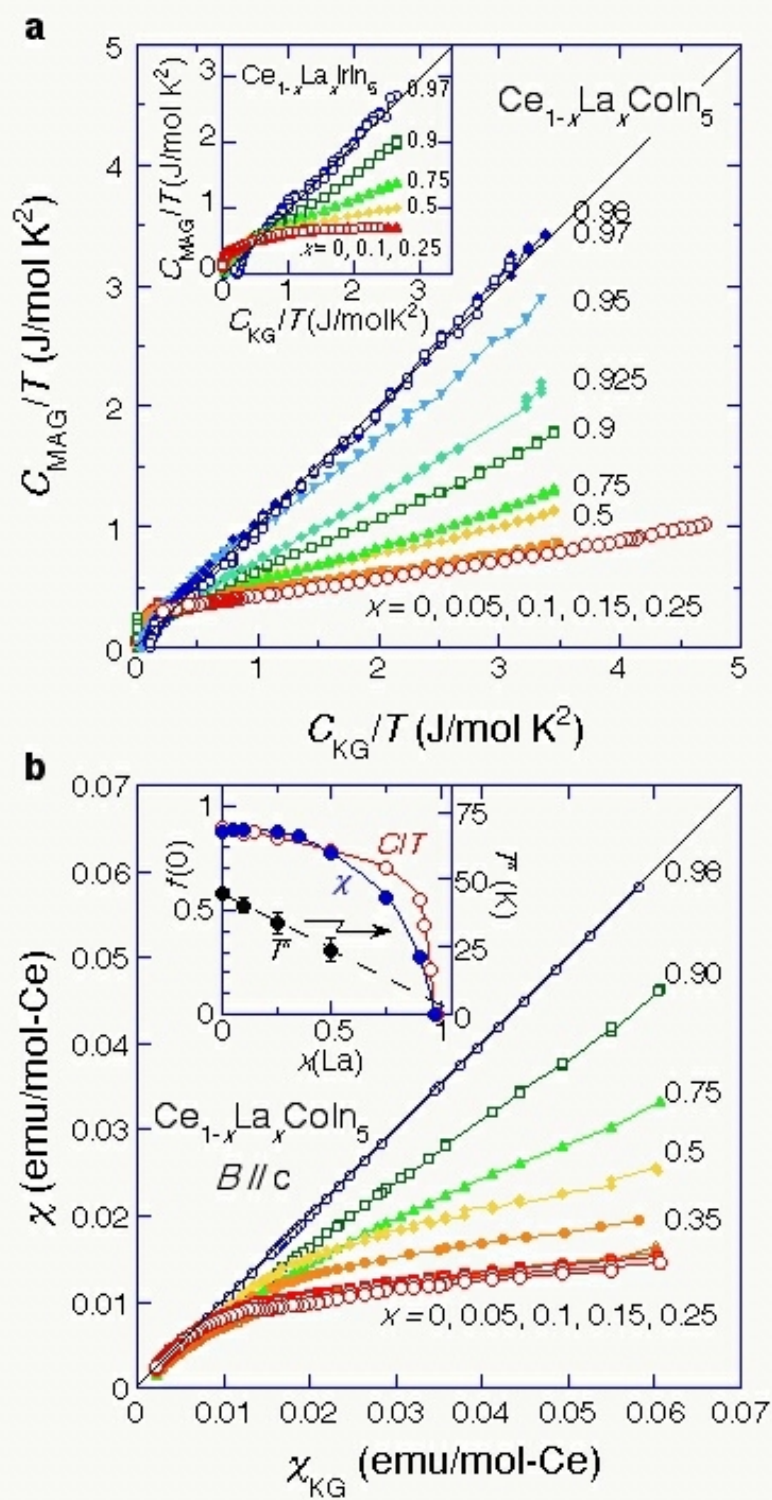

Figure 2

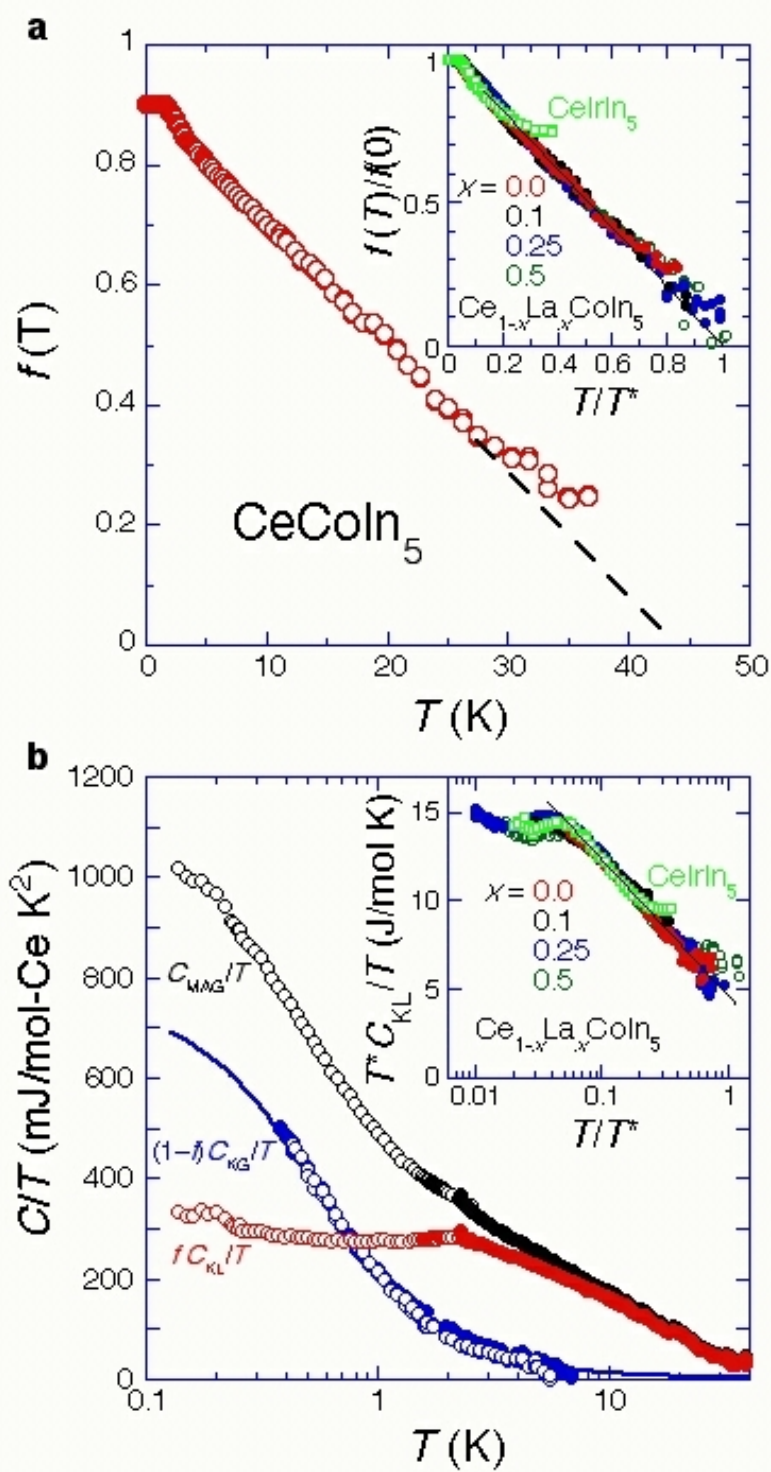

Figure 3

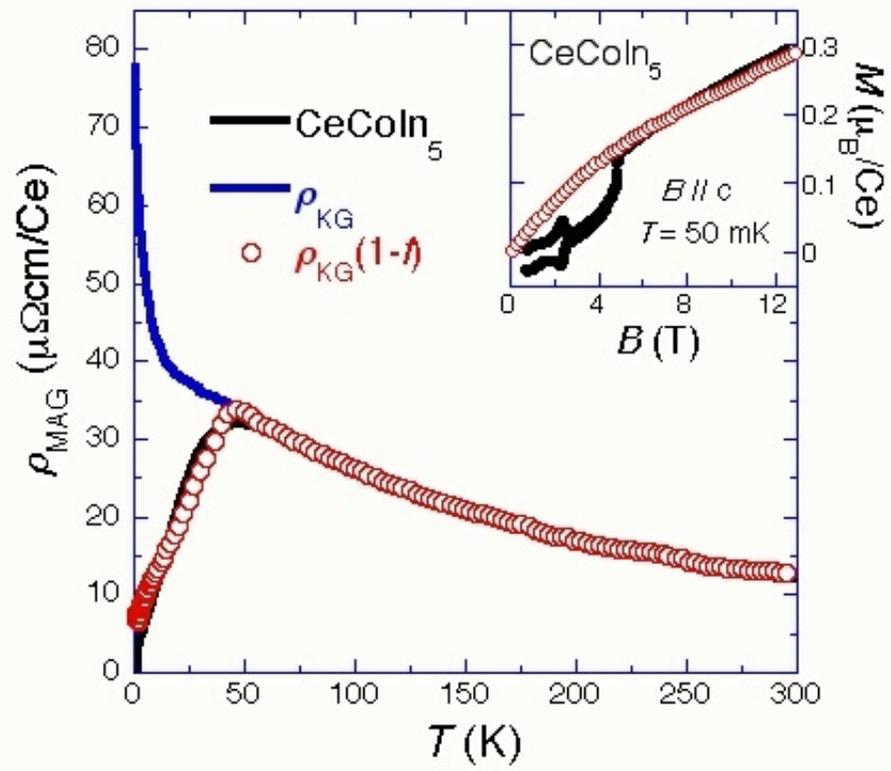

Figure 4